\documentclass[10pt, conference, letterpaper]{IEEEtran}
\IEEEoverridecommandlockouts

\usepackage{cite}
\usepackage{amsmath,amssymb,amsfonts,amsthm}
\usepackage{algorithmic}
\usepackage{graphicx}
\usepackage{textcomp}
\usepackage{xcolor}
\usepackage{graphicx}
\usepackage{subfigure}

\newtheorem{claim}{Claim}
\newtheorem{remark}{Remark}
\newtheorem{theorem}{Theorem}
\newtheorem{corollary}{Corollary}

\begin{document}
\newtheorem{defn}{Definition}
\title{Non-clairvoyant Scheduling of Coflows\thanks{Support of the DAE, Govt. of India, under project no. 12-R\&D-TFR-5.01-0500 and MATRICS grant by SERB India to Rahul Vaze is acknowledged.}}

\author{\IEEEauthorblockN{Akhil Bhimaraju}
\IEEEauthorblockA{\textit{IIT Madras}}
\and
\IEEEauthorblockN{Debanuj Nayak}
\IEEEauthorblockA{\textit{IIT Gandhinagar}}
\and
\IEEEauthorblockN{Rahul Vaze}
\IEEEauthorblockA{\textit{TIFR, Mumbai}}
}

\maketitle
\begin{abstract}
The coflow scheduling problem is considered: given an input/output switch with each port having a fixed capacity, find a scheduling algorithm that minimizes the weighted sum of the coflow completion times respecting the port capacities, where each flow of a coflow has a demand per input/output port, and coflow completion time is the finishing time of the last flow of the coflow. The objective of this paper is to present theoretical guarantees on approximating the sum of coflow completion time in the non-clairvoyant setting, where on a coflow arrival, only the number of flows, and their input-output port is revealed, while the critical demand volumes for each flow on the respective input-output port is unknown. The main result of this paper is to show that the proposed BlindFlow algorithm is $8p$-approximate, where $p$ is the largest number of input-output port pairs that a coflow uses. This result holds even in the online case, where coflows arrive over time and the scheduler has to use only causal information. Simulations reveal that the experimental performance of BlindFlow is far better than the theoretical guarantee. 
\end{abstract}
\section{Introduction}
\label{sec:intro}

Coflow scheduling is a recent popular networking
abstraction introduced to capture application-level computation and communication
patterns in data centers. For example, in distributed/parallel processing systems such as MapReduce \cite{dean2008mapreduce} or Hadoop \cite{shvachko2010hadoop}, Dryad \cite{isard2007dryad}, jobs/flows alternate between computation and communication stages, where a new stage cannot start until all the required sub-jobs/flows have been processed in the preceding stage. 
Therefore, the metric of performance is the delay seen by the last finishing job/flow in a stage unlike the conventional latency notion of per-job/flow delay. 

To better abstract this idea, a {\it coflow} \cite{coflow} is defined that consists of a group of flows, where the group is identified by the computation requirements. The completion time of a coflow is defined to be the completion time of the flow that finishes last in the group. The main ingredients of the basic scheduling problem are as follows. There is a switch with $m_i$ input and $m_o$ output ports, and each port can process jobs at a certain maximum capacity. Coflows arrive over time, where each coflow has a certain weight (measures relative importance) and each flow of a coflow corresponds to a certain demand volume that has to be processed over a particular input-output port pair. 
Among the currently outstanding coflows, the scheduler's job is to assign processing rates for all the flows (on respective input-output ports), subject to ports' capacity constraints, with the objective of minimizing the weighted sum of the coflow completion time.

The value of the time-stamp at which the coflow ``completes'' is defined as the completion time.
If we subtract the coflow's release time from the completion time, we get the amount of time the coflow spent in the system, and this is called the {\it flow time}.
The problem of minimizing the weighted sum of flow times has been shown to be NP-hard 
to {\bf even approximate} within constant factors even when demand volumes of all the flows are known \cite{orderscheduling}.
Thus, similar to prior work on coflow scheduling, in this paper, we only consider the completion time problem, where coflows can be released over time (the online problem). Solving the completion time problem even in the online case has been considered quite extensively in literature \cite{twoapprox, sincronia, ngarg} (and references therein).

The coflow scheduling problem (CSP) for minimizing completion time is NP-hard, since a special case of this problem,  the concurrent open shop (COS) problem is NP-hard
(see \cite{varys} for definition of COS and reduction of COS to CSP). Because of the NP-hardness, the best hope of solving the CSP is to find tight approximations. 
For the COS problem, the best known approximation ratio is $2$ \cite{cos2approx} that is also known to be tight \cite{2inapprox}. 

Work in approximating the CSP
began with intuitive heuristic algorithms such as Varys \cite{varys}, Baraat \cite{baraat}, and Orchestra \cite{orchestra} that showed reasonable empirical performance, which was then followed by theoretical work that showed that a $5$-approximation is possible \cite{shafiee, sincronia}. 
However, no tight lower bounds better than $2$ are known yet. A randomized \(2\) approximation was proposed in \cite{twoapprox}.
A $12$ approximation was also derived in \cite{sincronia} for the online case.

Prior theoretical work on CSP only considered a clairvoyant setting \cite{shafiee,sincronia,twoapprox}, where as soon as a coflow arrives, the demand volumes for each of its flows per input-output port are also revealed, which can be used to find the schedule. 
In general, this may not always be possible as argued in \cite{aalo} for various cases, such as pipelining used between different stages of computation \cite{isard2007dryad, condie2010mapreduce, rossbach2013dandelion} or task failures/speculation \cite{zaharia2012resilient, isard2007dryad, dean2008mapreduce}. 
Recent research has shown that prior knowledge of flow sizes is
indeed not a plausible assumption in many cases, but it might be possible to estimate the volumes \cite{dukic}.

In this paper, we consider the more general {\it non-clairvoyant} setting for solving the CSP, where on a coflow arrival, only its weight (its relative importance), the number of flows,  and their corresponding input-output ports are revealed, while the demand volumes for each flow on the respective input-output port is unknown. Any flow departs from the system as soon as its demand volume is satisfied. The departure is then notified to the algorithm.

Non-clairvoyant model for CSP has been considered in \cite{aalo}, where a heuristic algorithm Aalo based on {\em Discretized Coflow Aware Least Attained Service} (D-CLAS) was used. 
This method was further refined in \cite{dhs} using statistical models for flow sizes.
The objective of this paper is to present theoretical guarantees on approximating the non-clairvoyant CSP. 
There is significant work on non-clairvoyant scheduling in the theoretical computer science literature, for example for makespan minimization \cite{brecht1997competitive, he2008provably}, average stretch \cite{becchetti2000scheduling}, flowtime \cite{kalyanasundaram2003minimizing, becchetti2004nonclairvoyant}, flowtime plus energy minimization with speed scaling \cite{gupta2010nonclairvoyantly, gupta2012scheduling}, or with precedence constraints \cite{ngarg}, but to the best of our knowledge not on the CSP.

For the non-clairvoyant CSP, we propose an algorithm BlindFlow, which is also online (does not need information about future coflows).
%
%
Let $p$ be the largest number of distinct input-output port pairs any coflow uses.
The {\bf main result} of this paper is to show that BlindFlow is $8p$ approximate, where the guarantee is with respect to clairvoyant offline optimal algorithm. This result holds even in the online case, i.e., when coflows arrive (arbitrarily) over time and the algorithm only has causal information. 
As a corollary of our result, we get that a modified BlindFlow algorithm is $4p$ approximate for the COS problem (that is a special case of CSP) in the non-clairvoyant setting, which to the best of our knowledge was not known.  

Our proof technique involves expressing the clairvoyant {\it fractional} CSP  as a linear program (LP) and considering its dual. Then via the primal-dual method, we couple the rates allocated by BlindFlow to the dual variables of the clairvoyant fractional LP and then invoke weak duality, which is rather a novel idea in the CSP literature, inspired by \cite{ngarg}.

The approximation guarantees derived in this paper depend on the problem instance via the parameter $p$, 
and are not constant unlike the clairvoyant case \cite{shafiee, sincronia}. The reasons thereof are briefly commented on in Remark \ref{rem:nontrivial}.
The proof ideas are, however, novel, and the bounds are useful when the maximum number of ports each coflows uses, $p$, is small, or the number of total port pairs is small.  Moreover, as the simulations show (both synthetic and real-world trace data based), the performance of BlindFlow is far superior than the $8p$ approximation guarantee. The simulation performance of the BlindFlow algorithm is similar to the heuristic algorithm Aalo, even though Aalo outperforms BlindFlow because of multiple specific additions in Aalo which are appealing but are difficult to analyze.

\section{System Model}
\begin{figure}
\centering
   
        \includegraphics[width=0.4\textwidth]{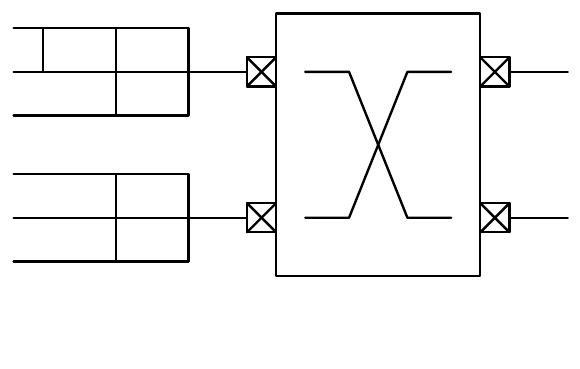}
       
\caption{A \(2 \times 2\) switch abstraction for datacenter networks, where 
the numbers within each flow indicate the coflow to which that
flow belongs.
Since we are dealing with the non-clairvoyant case, we do not 
know the sizes of the flows.
}

\label{fig:switch}
\end{figure}

Consider a switch with $m_i$ input and $m_o$ output ports as shown in Fig. \ref{fig:switch}. 

Without loss of generality we will assume that $m_i=m_o=m$.
{\it Coflow} \(k\) is the pair \((C_k, R_k)\), where \(C_k=[d_{ijk}]\) is an \(m \times m\)
matrix with non-negative entries and \(R_k\) is a non-negative real number, 
that represents its {\em release time}, the time after which it can be processed.
For the \((i,j)^\text{th}\) {\it flow} in coflow \(k\), 
we need to transfer \(d_{ijk}\) amount of data
from the \(i^\text{th}\) input port to the \(j^\text{th}\) 
output port of an \(m \times m\) switch. 

All the ports are capacity constrained and the \(i^\text{th}\) input port
can process \(c^\textsc{ip}_i\) units of data per unit time while the \(j^\text{th}\)
output port can process \(c^\textsc{op}_j\) units of data per unit time.
There are \(n\) coflows in the system \(\{(C_k, R_k)\}_{k=1}^n\), and
we want to schedule them in such a way to minimize the sum of weighted completion
times, as defined using the optimization program 
\ref{prog:opt} below.
\begin{align}
\underset{x_{ijkt},T_k \ge 0}{\text{minimize}}\quad&
    \sum\limits_k w_k T_k \tag{OPT} \label{prog:opt} \\
\text{subject to}\quad& 
    \sum\limits_t \frac{x_{ijkt}}{d_{ijk}} \ge 1 
        \quad \forall\ i,j,k, \label{eq:optcomplete} \\
& \sum\limits_{t > T_k} x_{ijkt} = 0
        \quad \forall\ i,j,k, \label{eq:defineTk} \\
\text{(for input port \(i\))}\quad& \sum\limits_{j,k} x_{ijkt} \le c^\textsc{ip}_i 
        \quad \forall\ i,t, \label{eq:optinputport}\\
\text{(for output port \(j\))}\quad& \sum\limits_{i,k} x_{ijkt} \le c^\textsc{op}_j 
        \quad \forall\ j,t, \label{eq:optoutputport}
\end{align}
where, \(\{w_k\}\)'s are the weights of each coflow, 
\(x_{ijkt}\) is the rate assigned to the \((i,j)^\text{th}\) flow of
coflow \(k\) at time \(t\), and $T_k$ is the completion time of coflow $k$. 
Constraints \eqref{eq:optcomplete} and \eqref{eq:defineTk} together ensure
that all the demands of a coflow are completed by time $T_k$. 
Constraints \eqref{eq:optinputport} and \eqref{eq:optoutputport} are capacity
constraints on input port \(i\) and output port \(j\) for \(i=1,2,..,m\) 
and \(j=1,2,..,m\).
Let \(\{x_{ijkt}^\textsc{opt},T_k^\textsc{opt}\}\) be the optimal solution 
to \ref{prog:opt} and let
the value of \ref{prog:opt} when we use these values be \(J_\textsc{opt}\),
i.e., \(J_\textsc{opt} = \sum_k w_k T_k^\textsc{opt}\).

\ref{prog:opt} cannot be solved except for small or trivial cases due to the
presence of \(T_k\) in the limits of summation in
constraint \eqref{eq:defineTk}.
The problem of finding the optimal schedule for minimizing weighted
completion time for coflows has been proven to be NP-hard \cite{varys}.
In prior work, assuming full knowledge of \(\{C_k\}\), algorithms with 
theoretical guarantees on their approximation ratios have been derived in \cite{shafiee, sincronia}. 

In this paper, as discussed in Sec. \ref{sec:intro}, we consider the 
non-clairvoyant case, where only the indices of the non-zero entries 
of \(\{C_k\}\) are revealed and not the exact values of \(\{d_{ijk}\}\). 
In addition, we assume that as the soon as the flow $k$ is released at time $R_k$, its weight is also revealed. 
This corresponds to only knowing the presence or absence of a flow to be 
fulfilled on a particular input-output pair but not the precise demand 
requirement on it. If weights are also {\bf unknown}, in the 
following, we can let all weights $w_k=1$ for all flows $k$ without changing any of our results.

Additionally, we consider the online setting, where we have no prior 
knowledge about the existence of a coflow before its release time \(R_k\).
Thus, at time \(t\), the information available is only about the set of flows 
that are yet to complete using the variables \(\mathbf{1}_{ijk}^t\) for \(k \in Q_t\),
where \(Q_t\) is the set of coflows released by time \(t\), i.e., 
\(Q_t = \{k \mid t \ge R_k\}\). 
\(\mathbf{1}_{ijk}^t\) is \(1\) if the \((i,j)^\text{th}\) flow of
coflow \(k\) is yet to finish and \(0\) otherwise.

Let \(n_{kt}=\sum_{i,j}\mathbf{1}_{ijk}^t\) be the number of 
unfinished flows of coflow \(k\) at time \(t\). 
Let \(\mathbf{1}_{k}^t\) be the indicator whether or not the entire coflow
has finished, i.e., \(\mathbf{1}_k^t=1\) if at least one among
\(\{\mathbf{1}_{ijk}^t \mid (i,j) \in m \times m \}\) is \(1\) and 
\(\mathbf{1}_k^t = 0\) otherwise.
When \(\mathbf{1}_k^t = 0\), let \(n_{kt}\) be any non-zero real number
to make notation simpler (this would mean 
\(\mathbf{1}_k^t/n_{kt}\) is always defined).

Next, we define a problem instance parameter $p$, which will be used to express our approximation guarantee for the non-clairvoyant CSP. 

\begin{defn}\label{defn:p}
Let \(p\) be the maximum number of unfinished flows in 
any coflow at any time, i.e.,
\(p=\underset{k}{\max} \{n_{k0}\}\), 
the maximum value among \(\{n_{kt}\}\) at \(t=0\). 
\end{defn}

Note that for a particular input-output port pair, we can have at most one flow per coflow (definition of coflow).
This implies that \(p\) is at most the maximum number of distinct input/output port pairs used by any coflow, and hence $p\le m^2$.

\section{BlindFlow Algorithm}

We propose the following  non-clairvoyant algorithm to 
approximate the problem \ref{prog:opt}.
We divide the capacity of a port among all the flows that
require that particular port in proportion to the flow weights.
This is a natural choice, since the demand volumes for each flow are unknown.
More precisely, BlindFlow allocates the rate \(r_{ijk}(t)\) in \eqref{eq:ratealloc} to the \(k^\text{th}\) coflow
on the \((i,j)^\text{th}\) input-output port pair at time \(t\), as
\vspace{-0.1in}
\begin{align}
r_{ijk}(t) = \frac{w_k \mathbf{1}_{ijk}^t}{
    \sum\limits_{l \in Q_t}\sum\limits_u \frac{w_l}{c^\textsc{op}_j} \mathbf{1}_{ujl}^t
    +  \sum\limits_{l \in Q_t}\sum\limits_v \frac{w_l}{c^\textsc{ip}_i} \mathbf{1}_{ivl}^t
}.
\label{eq:ratealloc}
\end{align}

For \(t < R_k\), \(r_{ijk}(t)\) is obviously \(0\).
Letting the ``weight'' of a flow to mean the weight of the coflow
it belongs to, an outstanding flow on input-output port pair $i,j$ gets a rate 
proportional to the ratio of its weight
and the sum of the weights of all other flows that need either the input
port \(i\) or the output \(j\), normalized to the port capacities \(c^\textsc{ip}_i\)
and \(c^\textsc{op}_j\).
Note that the flows that need both the input port \(i\) and the
output port \(j\) are counted twice in the denominator of \eqref{eq:ratealloc}.

\begin{remark}
Equation \eqref{eq:ratealloc} might produce a schedule where the rates of some
flows can be increased without violating the feasibility on any port.
A better rate allocation is
given by 
\begin{align*}
r_{ijk}(t) = \frac{w_k \mathbf{1}_{ijk}^t}{
\max (    \sum\limits_{l \in Q_t}\sum\limits_u \frac{w_l}{c^\textsc{op}_j} \mathbf{1}_{ujl}^t
\    ,\  \sum\limits_{l \in Q_t}\sum\limits_v \frac{w_l}{c^\textsc{ip}_i} \mathbf{1}_{ivl}^t )
},
\end{align*}
replacing the \(+\) operator with \(\max\) in the denominator.
Any performance guarantee on \eqref{eq:ratealloc} automatically holds for this rate
allocation as well.
\end{remark}

BlindFlow is a very simple algorithm, and is clearly non-clairvoyant
(does not use demand information $d_{ijk}$) and online (does not use information
about future coflow arrivals to schedule at the current time).

\textbf{An example}
Consider a simple example where we have a \(2 \times 2\) switch with 
port capacities \(1\) on all the ports and \(2\) 
coflows in the system. At some time \(t\), let the indicator matrices that
indicate the outstanding flows for these 
coflows be \(\mathbf{1}_1\) and \(\mathbf{1}_2\), where \(\mathbf{1}_1\) 
is the \(2 \times 2\) matrix \([\mathbf{1}_{ij1}^t]\) and \(\mathbf{1}_2\) is
defined similarly. 
For this example, assume that these indicator matrices are given by:
\begin{align*}
\mathbf{1}_1 = 
\begin{bmatrix}
1 & 0 \\
1 & 0
\end{bmatrix} \quad \text{and} \quad
\mathbf{1}_2 = 
\begin{bmatrix}
1 & 1 \\
0 & 1
\end{bmatrix}.
\end{align*}
This example is the same as the one shown in Fig. \ref{fig:switch}.
Let the weights for the coflows be \(w_1 = 1\) and \(w_2 = 2\).
From \eqref{eq:ratealloc}, the rate for the \((1,1)\) flow of coflow \(1\) is
\begin{align*}
r_{111} = \frac{1}{(1\cdot1+1\cdot1+2\cdot1)+(1\cdot1+2\cdot1+2\cdot1)}
= \frac{1}{9}.
\end{align*}
Similarly we get from \eqref{eq:ratealloc} the other rates as
\begin{align*}
r_1 = 
\begin{bmatrix}
1/9 & 0 \\
1/6 & 0
\end{bmatrix} \quad \text{and} \quad
r_2 = 
\begin{bmatrix}
2/9 & 2/9 \\
0 & 2/7
\end{bmatrix}.
\end{align*}
Here, the \(i,j\) entry of \(r_1\) is the rate allocated to the \((i,j)\)
port-pair of coflow \(1\). \(r_2\) is defined similarly.

The main result of this paper on the approximation ratio achieved by BlindFlow is as follows. 
\begin{theorem}\label{thm:main}
The rate allocation \eqref{eq:ratealloc} of the BlindFlow algorithm is feasible and is $8p$-approximate. In particular, if $J_\textsc{opt}$ is the optimal weighted coflow completion time,  then BlindFlow produces a schedule with a weighted coflow completion time 
that is no larger than \(8p \times J_\textsc{opt}\).
\end{theorem}
\begin{remark}
The approximation ratio guarantee is independent of the number of coflows, the volumes of coflows, and the capacities of the input-output ports, and is only a function of the parameter $p$ (Definition \ref{defn:p}). The parameter $p$, is the  number of flows with distinct input/output port requirements maximized over all co-flows. Theoretically, $p$ can be as large as $m^2$, however, for large switches (where $m^2$ is very large), 
actual coflows typically have \(p\) much smaller than $m^2$. Thus, the guarantee is still meaningful. Moreover, note that the approximation guarantee is with respect to the clairvoyant offline optimal algorithm. 
In the clairvoyant case, the approximation ratio guarantee (of $5$) is independent of the input \cite{shafiee,sincronia}.
However, introducing additional complexity into the problem, such as 
non-clairvoyance (this paper) or dependence across coflows \cite{2mplus1}, seems to
inevitably make the guarantee a function of \(m\).

\end{remark}

{\it Proof Sketch:} We prove Theorem \ref{thm:main} using a series of claims in the subsequent sections, where 
the main steps are as follows.
We first decrease the rates allocated by BlindFlow to a baseline rate.
Since these rates are lower, any guarantees on the baseline rate algorithm
apply to BlindFlow as well.
Then we ``speed-up'' the switch by a factor of \(4p\) while using
the baseline rates.
So any guarantee on the faster switch will apply to the original problem
with additional factor of \(4p\).
Next, we write a fractional LP formulation, \ref{prog:flp}, the value of whose optimal
solution is smaller than the optimal value of the objective we are
trying to minimize, \(J_\textsc{opt}\).
This implies that any dual feasible solution to \ref{prog:flp} will have a 
dual objective that is smaller than \(J_\textsc{opt}\).
We then produce a dual feasible solution using the speed-up rates whose dual objective is equal 
to half the weighted coflow completion time obtained by running the faster switch.
This gives us the \(8p\) guarantee after combining the \(4p\) penalty.

\begin{claim}
The rate allocation made by BlindFlow in \eqref{eq:ratealloc}  is always feasible.
\end{claim}

\begin{IEEEproof}
Any rate allocation that satisfies 
\(\sum\limits_{k \in Q_t}\sum\limits_{j}r_{ijk}(t) \le c^\textsc{ip}_i\) for
all input ports \(i\) and 
\(\sum\limits_{k \in Q_t}\sum\limits_{i}r_{ijk}(t) \le c^\textsc{op}_j\) for all
output ports \(j\) is a feasible schedule.
For any input port \(i\), we have,
\begin{align*}
\sum_{k \in Q_t}\sum_{j}r_{ijk}(t) &= 
    \sum_{k \in Q_t} \sum_j \frac{w_k \mathbf{1}_{ijk}^t}{
        \sum\limits_{l \in Q_t}\sum\limits_u \frac{w_l}{c^\textsc{op}_j} \mathbf{1}_{ujl}^t
        +  \sum\limits_{l \in Q_t}\sum\limits_v \frac{w_l}{c^\textsc{ip}_i} \mathbf{1}_{ivl}^t
    }, \\ 
    &\le
    \sum_{k \in Q_t} \sum_j \frac{w_k \mathbf{1}_{ijk}^t}{
        \sum\limits_{l \in Q_t}\sum\limits_v \frac{w_l}{c^\textsc{ip}_i} \mathbf{1}_{ivl}^t
    }
    = c^\textsc{ip}_i.
\end{align*}
Similar argument follows for each output port as well.
\end{IEEEproof}

Let coflow \(k\) finish at time \(T_k^\textsc{alg}\) 
when we use the schedule determined by \eqref{eq:ratealloc}.
Let \(J_\textsc{alg}\) be the weighted completion time produced
by BlindFlow, i.e., \(J_\textsc{alg} = \sum_k w_k T_k^\textsc{alg}\).

{\bf A baseline allocation}
To analyse the rates allocated by BlindFlow in \eqref{eq:ratealloc}, we define the following
``baseline'' algorithm for scheduling the coflows:
\begin{align}\label{eq:basealloc}
r_{ijk}^\textsc{base}(t) &= \begin{cases}
    \frac{w_k \mathbf{1}_{ijk}^t}{
        \sum\limits_l \sum\limits_u \frac{w_l}{c^\textsc{op}_j} \mathbf{1}_{ujl}^t
        + \sum\limits_l \sum\limits_v \frac{w_l}{c^\textsc{ip}_i} \mathbf{1}_{ivl}^t
    } 
    &  \text{for} \ t \ge 4p R_k  \\
 0 & \text{for} \  t < 4p R_k, 
\end{cases}
\end{align}
where $R_k$ is the release time of the coflow $k$.

Note that we do not require the rate allocation using the baseline algorithm 
\eqref{eq:basealloc} to be feasible or causal since we would not
actually run this algorithm on a switch.
We use the rates in \eqref{eq:basealloc} just as a means to upper bound the 
weighted completion time using \eqref{eq:ratealloc}, as we
show in the subsequent discussion.

Compared to \eqref{eq:ratealloc}, in \eqref{eq:basealloc}, we compute the rate using weights of all the unfinished flows, and
not just the ones that have been released.
In particular, the summation in the denominator here 
includes the flows that may be released in the future unlike \eqref{eq:ratealloc}.
Moreover, we do not schedule any flow in coflow \(k\) until time \(4p R_k\),
unlike BlindFlow, where we start scheduling as soon as it is released at \(t = R_k\).
The rate allocation to a particular flow using BlindFlow might decrease 
over time if new coflows are released.
This does not happen with the baseline rate allocation as we consider all
the unfinished coflows in the denominator of the expression.
However, the allocation in \eqref{eq:basealloc} gives us strictly smaller rates than what is 
allocated by BlindFlow because the denominator in BlindFlow can never be
greater than the sum of all the current {\em and} future flows sharing 
the same input or output port.
But as we see, the rates in \eqref{eq:basealloc} are sufficient to prove our performance guarantee.
Let the weighted completion time obtained by using the rates
allocated by \eqref{eq:basealloc} be \(J_\textsc{base}\).
\begin{claim}
\(J_\textsc{alg} \le J_\textsc{base}\).
\label{claim:alglebase}
\end{claim}
\begin{IEEEproof}
At any time \(t\), given the same set of unfinished flows, 
the rates we get by using \eqref{eq:ratealloc} are greater than 
or equal to the rates we get using \eqref{eq:basealloc} for every flow.
By using induction from \(t=0\), where the set of unfinished flows
would be the same for both the algorithms, we can conclude the claim.
\end{IEEEproof}

\section{The augmented switch}

Using ideas from \cite{ngarg}, we first prove an approximation guarantee
after ``speeding up'' the switch by a certain factor.
Later, we can relax this assumption at a cost to our guarantee 
equal to the speed-up factor.
For a switch, the speed-up is in terms of adding additional capacity
to its ports.
We now describe this setup formally.

Consider a switch where input ports have a capacity of \(4p\times c^\textsc{ip}_i\) 
instead of \(c^\textsc{ip}_i\) (likewise  for output ports).
This means that now we can process up to \(4p c^\textsc{ip}_i\) units of demand over each
port per time unit.
Hypothetically, consider scheduling the coflows over this
new faster switch using the following \(\{r_{ijk}^\textsc{aug}(t)\}\):
\vspace{-0.1in}
\begin{align}
r_{ijk}^\textsc{aug}(t) &= 4p \times 
    \frac{w_k \mathbf{1}_{ijk}^t}{
        \sum\limits_l \sum\limits_u \frac{w_l}{c^\textsc{op}_j} \mathbf{1}_{ujl}^t
        + \sum\limits_l \sum\limits_v \frac{w_l}{c^\textsc{ip}_i} \mathbf{1}_{ivl}^t
    }
    \label{eq:augalloc}
\end{align}
for  $t \ge R_k$.
Note that just like in \eqref{eq:basealloc}, we add the weights from all the
coflows in the denominator of \eqref{eq:augalloc} and not just the released
coflows like \eqref{eq:ratealloc}.
However, we start processing coflow $k$ at rate \eqref{eq:augalloc} at time \(R_k\) unlike \eqref{eq:basealloc},
where we wait till time \(4p\,R_k\).

Let the weighted completion time we obtain by running \eqref{eq:augalloc}
be \(J_\textsc{aug}\).
If we stretch the time axis by \(4p\) and reduce \(r_{ijk}^\textsc{aug}\)
by a factor of \(4p\), we get \(r_{ijk}^\textsc{base}\).
But since we are stretching the time axis, the completion times we get
by using \eqref{eq:basealloc} are \(4p\) times as big compared to the 
completion times produced by \eqref{eq:augalloc}.
This gives us the following claim.
\begin{claim}
\(J_\textsc{base} = 4p \times J_\textsc{aug}\),
where \(J_\textsc{aug}\) is the weighted sum of completion times
when we run the augmented rates \eqref{eq:augalloc}.
\label{claim:baseeq4paug}
\end{claim}

\begin{remark}
For the duration in which \(\mathbf{1}_{ijk}^t=1\), 
i.e., till the time the demand on the \((i,j)\) port pair
of coflow \(k\) is not fully satisfied,
the corresponding rate 
from \eqref{eq:augalloc}, \(r_{ijk}^\textsc{aug}(t)\),
is a non-decreasing function of \(t\).
This is because the terms in the denominator of \eqref{eq:augalloc} can only decrease as time
progresses.
\label{rem:ratesnondecreasing}
\end{remark}

From here on, \(\mathbf{1}_{ijk}^t\) would be the indicator of whether
or not the demand \(d_{ijk}\) has been satisfied {\em when we use the
rate allocation from \eqref{eq:augalloc}}.
In the following sections, we shall prove that using the rates from 
\eqref{eq:augalloc}, we get the sum of weighted completion times to be
no worse than twice what we get by using the optimal schedule for \ref{prog:opt}.
Since this is obtained by compressing the time axis by \(4p\), and the
rate allocation in \eqref{eq:augalloc} does
no worse than twice the optimal for \ref{prog:opt},
this gives us
the \(8p\) guarantee.

\section{The fractional LP}

Since the problem of minimizing weighted completion time (\ref{prog:opt}) is 
NP-hard, we use a simpler problem that can be written
as a linear program.
This is the problem of minimizing the ``fractional'' completion time.
For a single flow, the fractional completion time is calculated by
dividing the job into small chunks and taking the average completion time
of the different chunks.
Intuitively, fractional completion time should be less than the actual
completion time, since for the actual completion time we only consider the
time when the last chunk finishes.
We extend this concept to coflows and formally prove that fractional 
completion time is indeed less than the actual completion time.
This gives us a lower bound on \(J_\textsc{opt}\),
and as we see subsequently, \(J_\textsc{aug}\) is not far from this
lower bound.

Consider the following linear program that represents the fractional CSP.
\vspace{-0.1in}
\begin{align}
\underset{f_{kt},x_{ijkt}\ge0}{\text{minimize}}\quad & 
    \sum\limits_k w_k \sum\limits_{t \ge R_k} tf_{kt} \tag{FLP} 
        \label{prog:flp}\\ 
\text{subj. to}\quad & 
    \sum\limits_{s=R_k}^t f_{ks} 
        \le \sum\limits_{s=R_k}^t \frac{x_{ijks}}{
        d_{ijk}}, \tag*{}\ \  \forall\ i,j,k\ \text{with}\ t \ge R_k,
            \\  \label{eq:flpminfrac}\\
& \sum\limits_{t \ge R_k} f_{kt} \ge 1, 
        \quad \forall\ k, \label{eq:flpco-flowcomplete} \\
& \sum\limits_{k \in Q_t}\sum\limits_{i}x_{ijkt} \le c^\textsc{op}_j,
    \quad \forall\ j,t, \label{eq:flpoutputport} \\
& \sum\limits_{k \in Q_t}\sum\limits_j x_{ijkt} \le c^\textsc{ip}_i.
    \quad \forall\ i,t. \label{eq:flpinputport}
\end{align}
Here, \(\sum_{t \ge R_k} tf_{kt}\) is the 
fractional completion time of coflow \(k\). 
The variables \(\{x_{ijkt}\}\) and \(\{f_{kt}\}\) are 
defined for \(t \ge R_k\). 
Additionally, \(x_{ijkt}\) is only defined if \(d_{ijk} \neq 0\), but 
for simplicity of presentation, we drop mentioning this everywhere.
\(x_{ijkt}\) represents the rate allocated for the coflow $k$ on port pair $(i,j)$ at time $t$.
The fraction of demand \(d_{ijk}\) that has completed by  time \(t\) is given
by \(\sum_{s=R_k}^t\left(x_{ijks}/d_{ijk}\right)\).
We define \(f_{kt}\) so that \(\sum_{s=R_k}^t f_{ks}\) is equal
to the minimum among these \(\{\sum_{s=R_k}^t\left(x_{ijks}/d_{ijk}\right)\}\) 
fractions over all the flows of coflow \(k\).
Intuitively, \(f_{kt}\) is the ``fraction'' of coflow \(k\) that
has finished during time slot \(t\), so that \(\sum_{s=R_k}^t f_{ks}\) is
the fraction of coflow completed by time \(t\).

Constraint \eqref{eq:flpminfrac} defines \(\{f_{kt}\}\), and
constraint \eqref{eq:flpco-flowcomplete} ensures that the demands of all the flows in a coflow
have been completely satisfied eventually.
Constraints \eqref{eq:flpoutputport} and \eqref{eq:flpinputport} are similar
to constraints \eqref{eq:optoutputport} and \eqref{eq:optinputport} and
ensure that capacity constraints are satisfied for all the ports.

The following is a formal proof of the above intuitive ideas 
that \ref{prog:flp} is indeed a lower bound for \ref{prog:opt}.

\begin{claim}
The optimal value of \ref{prog:flp} is a lower bound on
\(J_\textsc{opt}\), the optimal value of \ref{prog:opt}.
\label{claim:flplowerbound}
\end{claim}
\begin{IEEEproof}
Consider the optimal schedule \(\{x_{ijkt}^\textsc{opt}\}\) of
\ref{prog:opt}.
Note that this need not be the optimal solution for \ref{prog:flp}.

Since \(\{x_{ijkt}^\textsc{opt}\}\) is a feasible schedule for \ref{prog:opt}, the 
capacity constraints \eqref{eq:flpinputport} and \eqref{eq:flpoutputport}
are satisfied as \eqref{eq:optinputport} and \eqref{eq:optoutputport} are
satisfied.
Define
\begin{align*}
f_{kt}^* = \underset{i,j}{\text{min}}
                \left\{
                    \sum\limits_{s=R_k}^t \frac{x_{ijks}^\textsc{opt}}{
                                                                    d_{ijk}}
                 \right\}
            - \underset{i,j}{\text{min}}
                \left\{
                    \sum\limits_{s=R_k}^{t-1} \frac{x_{ijks}^\textsc{opt}}{
                                                                    d_{ijk}}
                \right\},
\end{align*}
$\forall \ k,t \ge R_k$.
Since \(\sum\limits_{s=R_k}^t \frac{x_{ijks}}{d_{ijk}}\) (amount of satisfied demand) is a non-decreasing
function of \(t\ (\ge R_k)\) for any feasible schedule \(\{x_{ijkt}\}\),
for any \((i,j)\), the \(\min\) of all these 
functions over \(i,j\) is also a non-decreasing function.
This ensures that \(f_{kt}^* \ge 0\) for all \(k,t \ge R_k\).
From the definition of \(f_{kt}^*\), via a telescopic sum argument,
we have,
\begin{align}
\sum\limits_{s=R_k}^t f_{ks}^* & = \underset{i,j}{\text{min}}
                \left\{
                    \sum\limits_{s=R_k}^t \frac{x_{ijks}^\textsc{opt}}{
                                                                    d_{ijk}}
                \right\}             \le \sum\limits_{s=R_k}^t \frac{x_{ijks}^\textsc{opt}}{d_{ijk}},
                        \label{eq:flpconstr1}
\end{align}
$\forall \ i,j,k\ \text{with}\ t \ge R_k$.
Since \(\{x_{ijkt}^\textsc{opt}\}\) is a feasible schedule for \ref{prog:opt},  from constraint \eqref{eq:optcomplete} in \ref{prog:opt}, we get
$\sum\limits_{t \ge R_k}\frac{x_{ijkt}^\textsc{opt}}{d_{ijk}} = 1,
                                                \quad \forall \ i,j,k$.
This gives us
\vspace{-0.1in}
\begin{align}
\sum\limits_{t \ge R_k} f_{kt}^* = \underset{i,j}{\text{min}}
                \left\{
                    \sum\limits_{t \ge R_k} \frac{x_{ijkt}^\textsc{opt}}{
                                                                    d_{ijk}}
                \right\}
                    = 1 \quad \forall \ k.
                \label{eq:flpconstr2}
\end{align}
Equations \eqref{eq:flpconstr1} and \eqref{eq:flpconstr2} ensure the feasibility
of the solution \(\{x_{ijkt}^\textsc{opt}, f_{kt}^*\}\) for \ref{prog:flp}.
Now we show that \ref{prog:flp} objective for \(\{x_{ijkt}^\textsc{opt}, f_{kt}^*\}\)
is less than \(J_\textsc{opt}\)
From constraints \eqref{eq:optcomplete} and \eqref{eq:defineTk},
for all \(i,j\), we have
\begin{align*}
\sum\limits_{s=R_k}^t\frac{x_{ijks}^\textsc{opt}}{d_{ijk}} & = 1
    \quad \forall \ t \ge T_k^\textsc{opt},
    \implies f_{kt}^* & = 0 \quad \forall \ t > T_k^\textsc{opt}.
\end{align*}
Therefore we get
$
\sum\limits_{t \ge R_k} tf_{kt}^*
    \le \sum\limits_{t=R_k}^{T_k^\textsc{opt}} T_k^\textsc{opt}f_{kt}^*
    = T_k^\textsc{opt}.
$
This gives us
\begin{align*}
\sum\limits_k w_k \sum\limits_{t \ge R_k} tf_{kt}^*
    \le \sum\limits_k w_k T_k^\textsc{opt} = J_\textsc{opt}.
\end{align*}

As we have a feasible solution \(\{x_{ijkt}^\textsc{opt}, f_{kt}^*\}\),
where \ref{prog:flp} has a value less than
or equal to \(J_\textsc{opt}\), the optimal solution to \ref{prog:flp} 
will also have a value less than or equal to \(J_\textsc{opt}\).
\end{IEEEproof}

Next, we consider the dual program of \ref{prog:flp}, and show that the 
optimal dual objective is greater than or equal to half the sum of weighted
completion times, \(J_\textsc{aug}\), obtained by using the rates allocated by \eqref{eq:augalloc}.

\section{The dual program}

Let the dual variables corresponding to constraints
\eqref{eq:flpminfrac}, \eqref{eq:flpco-flowcomplete},
\eqref{eq:flpoutputport}, and \eqref{eq:flpinputport} be
\(\gamma_{ijkt}\), \(\alpha_k\), \(\phi_{jt}\), and \(\theta_{it}\)
respectively.
The dual program for \ref{prog:flp} is given by the following.
\begin{align}
\underset{\alpha,\phi,\theta,\gamma\ge0}{\text{maximize}}\quad & 
    \sum\limits_k \alpha_k - \sum\limits_{j,t}c^\textsc{op}_j\phi_{jt} - 
        \sum\limits_{i,t}c^\textsc{ip}_i\theta_{it} \tag{DLP} \label{prog:dlp}\\
\text{subject to}\quad & \alpha_k \le tw_k + \sum\limits_{i,j} 
    \sum\limits_{s\ge t}\gamma_{ijks} \quad \forall\ k,t \ge R_k,
        \label{eq:fkteq} \\
& \sum\limits_{s\ge t}\frac{\gamma_{ijks}}{d_{ijk}} \le \phi_{jt} 
        + \theta_{it} \quad \forall\ i,j,k,t \ge R_k. \label{eq:xijkteq}
\end{align}

We will define a new variable \(\alpha_{kt}\) and use this to set 
\(\alpha_k\).
Recall that we are using the rates allocated by \eqref{eq:augalloc},
and \(\mathbf{1}_{ijk}^t\) is \(1\) if the \((i,j)\) flow of
coflow \(k\) has not yet finished by time \(t\) with rates  \eqref{eq:augalloc} and \(0\) otherwise,
and \(n_{kt} = \sum_{i,j}\mathbf{1}_{ijk}^t\) is the total number of unfinished
flows in coflow \(k\) at time \(t\) when using rates \eqref{eq:augalloc}.
Consequently, define the dual variables as follows.
\begin{align}\label{defn:dualalpha}
\alpha_{kt} = w_k\mathbf{1}_{k}^t,\ \alpha_k = \sum\limits_t \alpha_{kt}, \\ \label{defn:dualtheta}
\theta_{it} = \frac{1}{4c^\textsc{ip}_i}
    \sum\limits_{j,k}\frac{w_k}{n_{kt}} \mathbf{1}_{ijk}^t, \\ \label{defn:dualphi}
\phi_{jt} = \frac{1}{4c^\textsc{op}_j}
    \sum\limits_{i,k}\frac{w_k}{n_{kt}} \mathbf{1}_{ijk}^t, \\\label{defn:gamma}
\gamma_{ijkt} = \frac{w_k}{n_{kt}} \mathbf{1}_{ijk}^t.
\end{align}

Note that the choice of the dual variables \eqref{defn:dualalpha}, 
\eqref{defn:dualtheta}, \eqref{defn:dualphi}, and  \eqref{defn:gamma}, 
and hence the dual objective, 
depends on the rates \eqref{eq:augalloc} via \(\{\mathbf{1}_{ijk}^t\}\) and \(\{n_{kt}\}\).
Let the value of \ref{prog:dlp}, when we process coflows with rates \eqref{eq:augalloc} 
on the augmented switch and define the dual variables as above,
be \(J_\textsc{dual}\).

\begin{claim}
\(J_\textsc{dual} = \frac{1}{2}J_\textsc{aug}\).
\label{claim:dualeqhalfaug}
\end{claim}

\begin{IEEEproof}
First consider the first term in the dual objective:
\begin{align}
\sum\limits_k \alpha_k = \sum\limits_{k,t} \alpha_{kt}
= \sum\limits_t \sum\limits_k w_k \mathbf{1}_k^t
= J_\textsc{aug}. \label{eq:alphacost}
\end{align}

The second term in the dual objective is:
\begin{align}
\sum\limits_{j,t}c^\textsc{op}_j \phi_{jt} &= \frac{1}{4}\sum\limits_{i,j,k,t}
            \frac{w_k}{n_{kt}}\mathbf{1}_{ijk}^t, \tag*{} \\
&= \frac{1}{4}\sum\limits_t \sum\limits_k w_k\left(\frac{1}{n_{kt}}
            \sum\limits_{i,j} \mathbf{1}_{ijk}^t\right), \tag*{} \\
&= \frac{1}{4}\sum\limits_t \sum\limits_k w_k \mathbf{1}_{k}^t
= \frac{1}{4} J_\textsc{aug}. \label{eq:phicost}
\end{align}

Similarly, we can show that
\begin{align}
\sum\limits_{i,t} \theta_{it}=\frac{1}{4}J_\textsc{aug}.\label{eq:thetacost}
\end{align}

Combining \eqref{eq:alphacost}, \eqref{eq:phicost}, and \eqref{eq:thetacost}
proves the claim.
\end{IEEEproof}

Next, we show that the dual  variables \eqref{defn:dualalpha}-\eqref{defn:gamma} are feasible for \ref{prog:dlp}.

\begin{claim}
The defined dual variables \eqref{defn:dualalpha}, \eqref{defn:dualtheta}, 
\eqref{defn:dualphi}, and  \eqref{defn:gamma} are feasible when running 
the augmented switch rates \eqref{eq:augalloc},
i.e., \(J_\textsc{dual}\) is produced by 
a feasible solution to \ref{prog:dlp}.
\label{claim:dualfeasible}
\end{claim}

\begin{IEEEproof}
For any \(t \ge R_k\), \(\alpha_k=\sum\limits_{s < t}\alpha_{ks} 
                            + \sum\limits_{s \ge t}\alpha_{ks}\).
Since \(\alpha_{ks}=w_k\mathbf{1}_k^s\),
\(\sum\limits_{s < t}\alpha_{ks} \le tw_k\).
From the definition of \(\gamma_{ijkt}\), we get
$\sum\limits_{i,j}\gamma_{ijks} 
    = \sum\limits_{i,j}\frac{w_k}{n_{ks}}\mathbf{1}_{ijk}^s
    = \frac{w_k}{n_{ks}}\sum\limits_{i,j}\mathbf{1}_{ijk}^s
    = w_k \mathbf{1}_k^s.$
Since \(\alpha_{ks}=\sum\limits_{i,j}\gamma_{ijks}\),
\(\sum\limits_{s \ge t}\alpha_{ks} \le 
    \sum\limits_{i,j}\sum\limits_{s \ge t}\gamma_{ijks}\)
holds with equality.
This shows the feasibility of constraint \eqref{eq:fkteq}.

Now we show the feasibility of constraint \eqref{eq:xijkteq}.
If \(\mathbf{1}_{ijk}^t = 0\), then the constraint is clearly satisfied.
Consider the case where \(\mathbf{1}_{ijk}^t = 1\).
Since \(n_{ks} \ge 1\) whenever \(\mathbf{1}_{ijk}^s = 1\),
\begin{align*}
\sum\limits_{s \ge t}\frac{\gamma_{ijks}}{d_{ijk}} 
    = \frac{1}{d_{ijk}}\sum\limits_{s \ge t}
                                \frac{w_k}{n_{ks}}\mathbf{1}_{ijk}^s
    \le \frac{w_k}{d_{ijk}}\sum\limits_{s \ge t}\mathbf{1}_{ijk}^s.
\end{align*}
The summation term is the extra time after \(t\) to finish the flow.
Since the rates \(\{r_{ijk}^\textsc{aug}(t)\}\) from \eqref{eq:augalloc} 
are non-decreasing with \(t\) for each flow (as long as the flow has 
not yet unfinished), and for any \(t \ge R_k\), the maximum 
amount of remaining data to be sent for the \((i,j)^\text{th}\) flow
of coflow \(k\) is \(d_{ijk}\), \(\sum_{s \ge t}\mathbf{1}_{ijk}^s\) 
is upper bounded by \(d_{ijk}/r_{ijk}^\textsc{aug}(t)\).
So we have,
\begin{align}
\sum\limits_{s \ge t}\frac{\gamma_{ijks}}{d_{ijk}} 
            \le \frac{w_k}{r_{ijk}^\textsc{aug}(t)}.
\label{eq:remtime}
\end{align}
Using \eqref{eq:augalloc}, we get
\begin{align*}
\sum\limits_{s \ge t}\frac{\gamma_{ijks}}{d_{ijk}} \le 
    \frac{1}{4p}\left( 
                    \sum\limits_{l,u}\frac{w_l}{c^\textsc{op}_j}\mathbf{1}_{ujl}^t
                    + \sum\limits_{l,v}\frac{w_l}{c^\textsc{ip}_i}\mathbf{1}_{ivl}^t
                \right).
\end{align*}

As \(p\) is the maximum number of flows that any coflow can have,
\(p \ge n_{kt}\) for any \(k\) and \(t\), and this gives us
\begin{align*}
\sum\limits_{s \ge t}\frac{\gamma_{ijks}}{d_{ijk}} &\le
    \sum\limits_{l,u}\frac{w_l}{4c^\textsc{op}_jn_{lt}}\mathbf{1}_{ujl}^t
    + \sum\limits_{l,v}\frac{w_l}{4c^\textsc{ip}_in_{lt}}\mathbf{1}_{ivl}^t 
= \phi_{jt} + \theta_{it}.
\end{align*}
\end{IEEEproof}
\vspace{-0.1in}
\begin{remark}\label{rem:nontrivial}
We can now highlight why we need a speed up factor of \(4p\) in 
\eqref{eq:augalloc}, which is the main reason our approximation
guarantee is a function of \(p\), and not a constant as in the clairvoyant 
case \cite{shafiee, sincronia}.
If we speed up by a constant factor, we cannot upper bound \eqref{eq:remtime}
by \(\phi_{jt}+\theta_{it}\), as \(n_{kt}\) might be large  for some coflow 
\(k\) (if we have a large number of unfinished flows),
and this would make \(\phi_{jt}+\theta_{it}\) too small.
Alternatively, we could try to incorporate \(n_{kt}\) into BlindFlow's rate
allocation \eqref{eq:ratealloc}, but this breaks the monotonicity of
\eqref{eq:augalloc} (Remark \ref{rem:ratesnondecreasing}), and we can
no longer be sure that \eqref{eq:remtime} holds. 
We would need to use
\(\min_{s \ge t}r^\textsc{aug}_{ijk}(s)\) instead, and
this would again lead to similar guarantees.
\end{remark}


Finally, we complete the proof of Theorem \ref{thm:main}.
\begin{IEEEproof}[Proof of Theorem \ref{thm:main}]
Since \(J_\textsc{dual}\) is produced by a feasible solution to the dual
of \ref{prog:flp} (from claim \ref{claim:dualfeasible}),
and the optimal solution to \ref{prog:flp} is a lower bound on 
\(J_\textsc{opt}\) (from claim \ref{claim:flplowerbound}),
we have
\vspace{-0.2in}
\begin{align*}
J_\textsc{dual} \le J_\textsc{opt}.
\end{align*}

Using claims \ref{claim:alglebase}, \ref{claim:baseeq4paug}, and 
\ref{claim:dualeqhalfaug}, we get
\begin{align*}
J_\textsc{alg} \le J_\textsc{base} 
                = 4p\ J_\textsc{aug} 
                = 8p\ J_\textsc{dual}
                \le 8p\ J_\textsc{opt}.
\end{align*}
\vspace{-0.1in}
\end{IEEEproof}
\vspace{-0.2in}

\subsection{Concurrent open shop problem}

The concurrent open shop problem is a special case of Problem \eqref{prog:opt} when all the 
matrices \(\{C_k\}\) are diagonal and the capacities are all 1 (see \cite{varys}).
In this case, we can improve the bound of $8p$ that we obtained as follows. 
While allocating rate \(r_{ijk}(t)\) for the flow from the input port
\(i\) to output port \(j\) of coflow \(k\) using BlindFlow in \eqref{eq:ratealloc}, 
if another flow shares both the input {\em and} output port with this flow, its weight is counted
two times in the denominator.
If all the matrices are diagonal, then any two flows that share either
the input or the output port necessarily have to share both the input
and the output port.
So every term in the denominator of \eqref{eq:ratealloc} is counted twice
when \(\{C_k\}\)'s  are all diagonal.
So we can double the rates in \eqref{eq:ratealloc} without violating
feasibility.
Formally, the rates \(\{r_{iik}^\textsc{conc}(t)\}\) to be used for 
the concurrent open shop problem are given by
\begin{align}
r_{iik}^\textsc{conc}(t) = \frac{w_k \mathbf{1}_{iik}^t}{
    \sum\limits_{l \in Q_t} w_l \mathbf{1}_{iil}^t
}.
\label{eq:concalloc}
\end{align}
Since the only flows are on port pairs \((i,j)\) with \(i=j\), we do not
need the other term in the denominator of \eqref{eq:ratealloc} to ensure
feasibility.

The baseline algorithm will now have double the rates 
in \eqref{eq:basealloc}. 
Let coflow  \(k\) now start from time \(2p\,R_k\) instead of \(4p\,R_k\) in
the baseline algorithm.
Since the rates for the baseline algorithm are twice those in \eqref{eq:basealloc},
and it now starts at \(2p\,R_k\) instead of \(4p\,R_k\), we can get to 
the same rates as in \eqref{eq:augalloc} with just a time
stretching of \(2p\) intead of \(4p\).
This gives us \(J_\textsc{base} = 2p\,J_\textsc{aug}\)
in claim \ref{claim:baseeq4paug} instead of \(4p\,J_\textsc{aug}\), 
and thus a \(4p\) approximation instead of the \(8p\) one.
This leads us to corollary \ref{cor:conc}.
\begin{corollary}
\label{cor:conc}
For the concurrent open shop problem, the rate allocation
\(\{r_{ijk}^\textsc{conc}(t)\}\) in \eqref{eq:concalloc} is
feasible and produces a schedule no worse than \(4p\) times the optimal.
\end{corollary}

\section{Experimental Results}

\begin{figure*}
\centering
    
        \includegraphics[width=.75\textwidth]{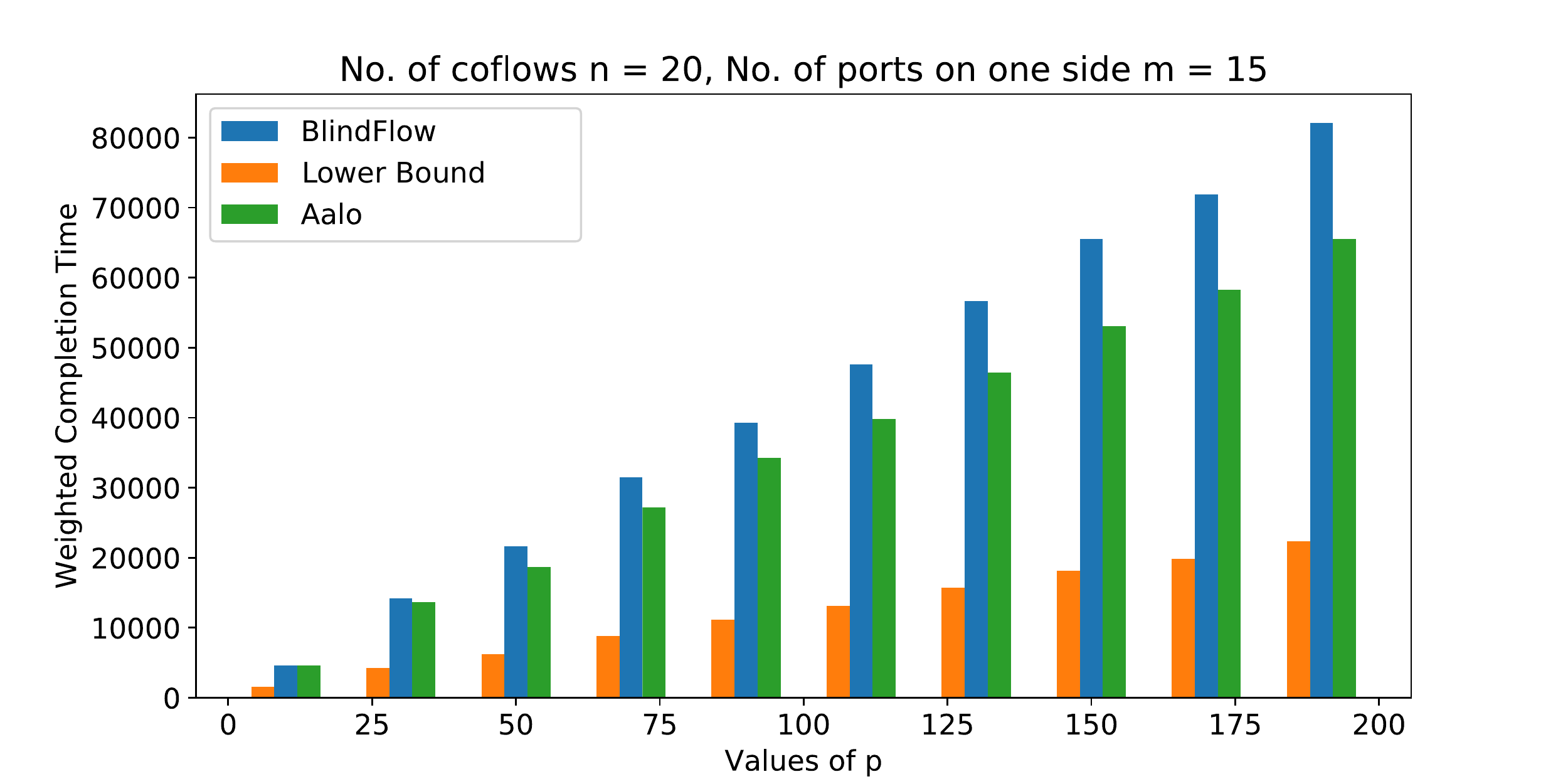}
        \label{fig:cp}
        \caption{Coflow completion times as a function of $p$.}
        \end{figure*}

In this section, we present some experimental results for the BlindFlow algorithm and compare it against a clairvoyant lower bound,
the relaxed LP from \cite{shafiee}, and the non-clairvoyant algorithm Aalo \cite{aalo}. 
We use two types of data to simulate their performance, synthetic data, and real world data from a facebook cluster. For generating the synthetic data we use the following procedure. 
\begin{enumerate}
    \item The number of coflows $n$, number of ports on each side $m$, the maximum number of non-zero entries in the demand matrices $p$, maximum demand for any flow $D$, and the last release time for any coflow $T$ are given as parameters.
    \item For each coflow $k$
    \begin{enumerate}
    \item a  number between $1$ and $p$ is chosen uniformly at random, which is the number of non zero entries in that coflow, defined as $p_k$. 
    \item $p_k$ many $(i,j)$ input-output pairs corresponding to the $p_k$ flows of coflow $k$ are chosen uniformly at random from the $m^2$ possible input-output port pairs.
    \item Each of $p_k$ pairs is given a demand from $1$ to $D$ chosen uniformly at random.
    \item For each coflow, a release time is chosen uniformly at random from a time interval from $[0,T]$.
    \end{enumerate}
\end{enumerate}
Using this synthetic data, we run the following experiments, that illustrate the effect of the parameter $p$ and the number of coflows $n$ on the performance of BlindFlow. In Fig. \ref{fig:cp} and \ref{fig:cn}, we compare the performance of BlindFlow, a clairvoyant lower bound on the coflow completion time from \cite{shafiee}, and the non-clairvoyant algorithm Aalo, as a function of $p$ and $n$ respectively.
For Fig. \ref{fig:cp}, we use $n = 20$, number of ports on each side $m= 15$, maximum demand on any flow $D = 15$ and last release time $T= 50$, while for Fig. \ref{fig:cn}, we use $p = 140$ and keep all the other parameters the same. The performance of the BlindFlow algorithm is close to but inferior to that of Aalo.
However, BlindFlow is much easier to implement than Aalo.
The performance of the non-clairvoyant BlindFlow is worse than the clairvoyant lower bound as expected.
Importantly, the ratio between the two does {\bf not} seem to scale with $p$, and is relatively small in contrast to the theoretical guarantee of $8p$ we have obtained. 
   

   \begin{figure*}
\centering     
        \includegraphics[width=.75\textwidth]{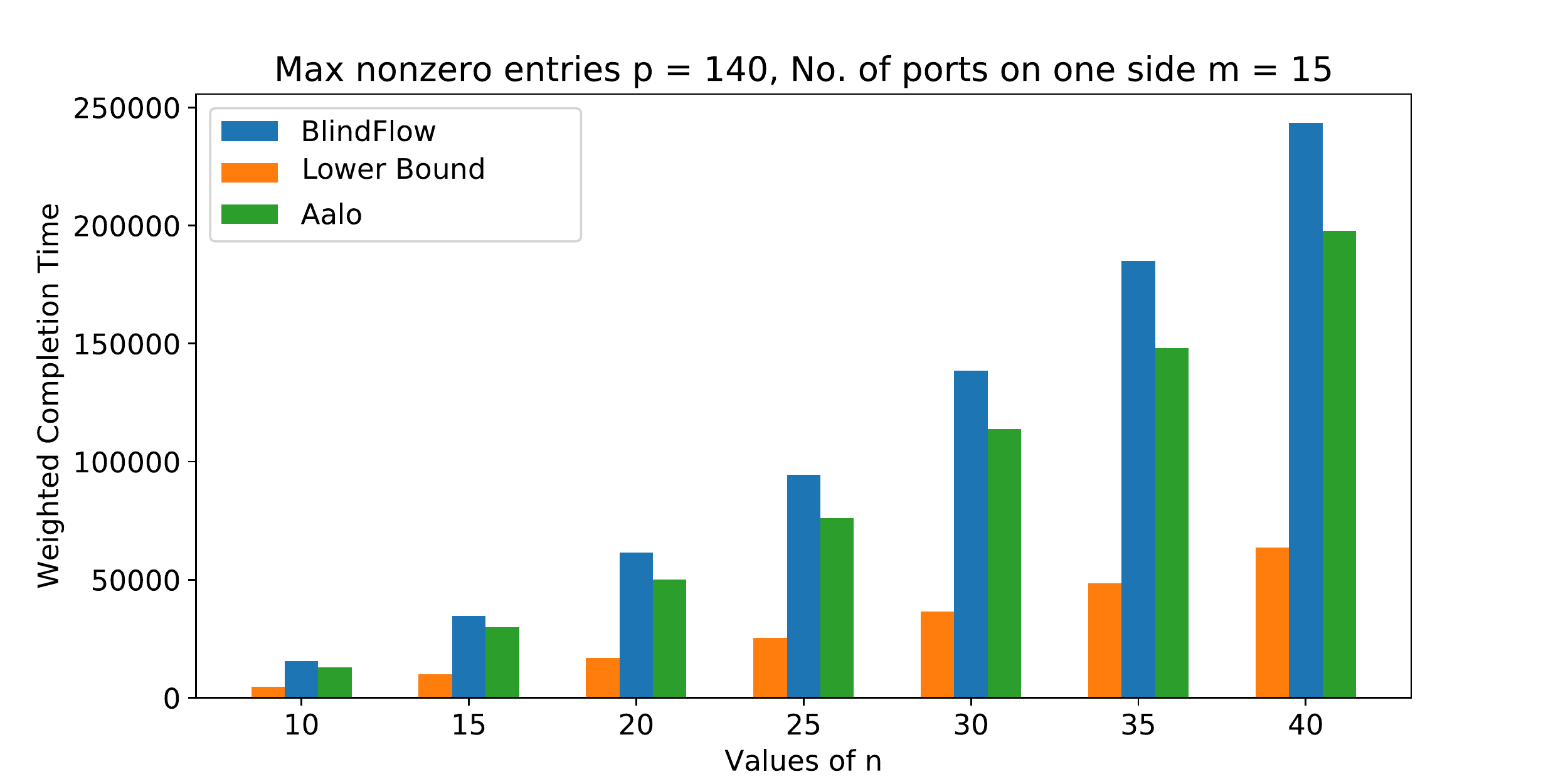}
        \label{fig:cn}
\caption{Coflow completion times as a function of $n$.}
\label{fig:cn}
\end{figure*}

\begin{figure*}
\centering
\includegraphics[width=.6\textwidth]{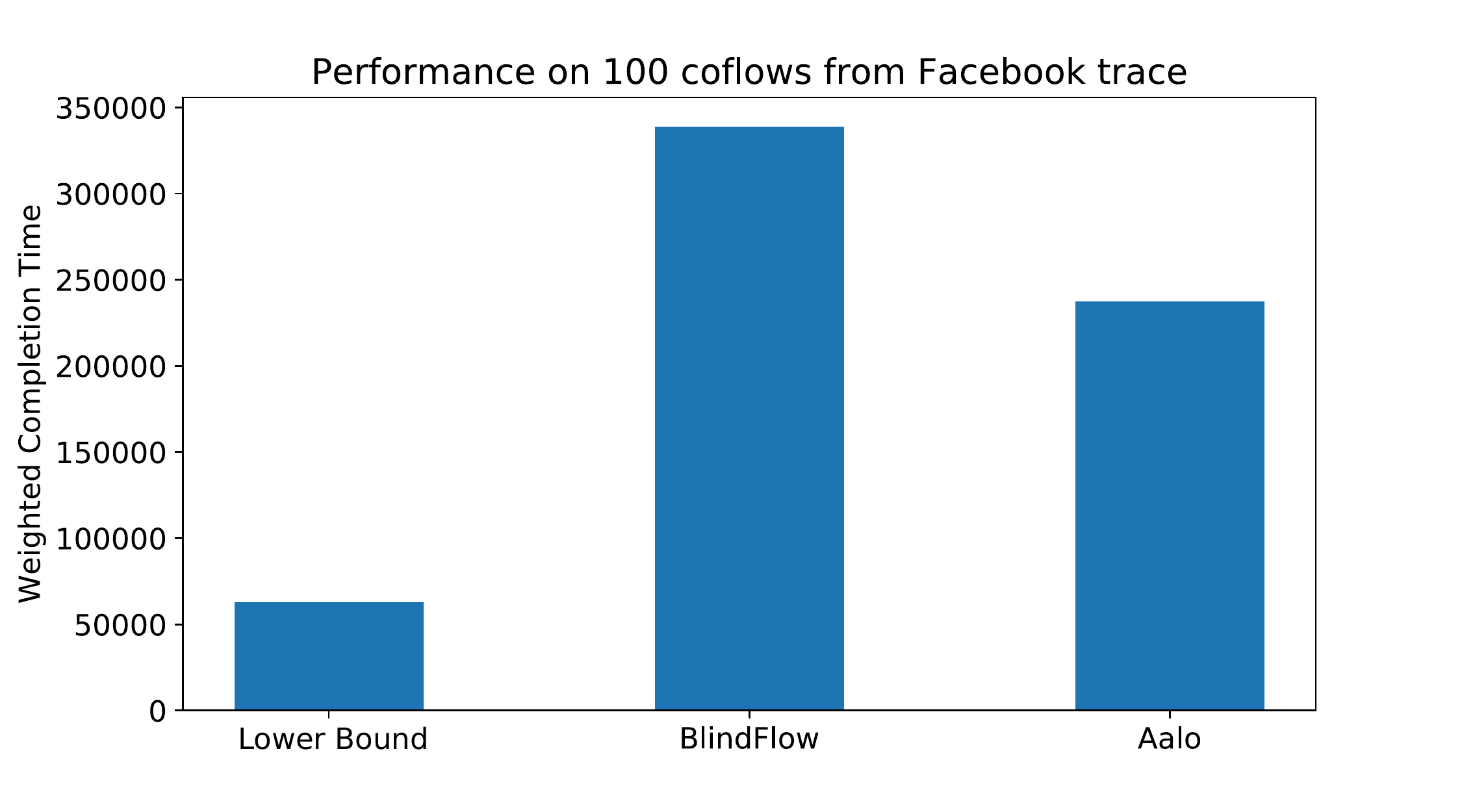}
\caption{Coflow completion times for the real world facebook trace data.}\label{fig:fb}
\end{figure*}
Next, in Fig. \ref{fig:fb}, we compare the performance of BlindFlow on the real world data that is  based on a Hive/MapReduce trace  collected by 
Chowdhury et  al. \cite{coflow} from a Facebook  cluster available at \cite{datalocation}. This trace has been used previously as well \cite{shafiee, varys, aalo}. The original trace is from a $3000$-machine 150-rack MapReduce cluster at Facebook. 
The original trace has $526$ coflows, however, for simulation feasibility on limited machines, we use the first $100$ coflows from this trace and execute the three algorithms on this. For our simulation, we assume that the rate of flow of any link at maximum capacity is $1$ MBps. Once again we see that the performance of BlindFlow is far better than the $8p$ guarantee that we have derived compared to the clairvoyant lower bound.
Moreover, for this simulation as well, 
Aalo outperforms BlindFlow, however, as stated before, BlindFlow is easier to implement, and is amenable for obtaining theoretical guarantee compared to the clairvoyant optimal algorithm, unlike Aalo for which no theoretical guarantee is available.

\section{Conclusions}
In this paper, for the first time, we have derived theoretical guarantees on the approximation ratio of weighted coflow completion time problem in the non-clairvoyant setting. The non-clairvoyant setting is both more robust, since the exact demand is unknown, and theoretically challenging, since we are comparing against the optimal algorithm that is clairvoyant. The guarantee we obtain compared to the clairvoyant optimal algorithm is a function of $p$, the maximum number of flows that any coflow can have, however, as shown via simulations, the actual performance is superior to the derived guarantee. It is not clear immediately whether the guarantee is a function of $p$, because we are comparing against the clairvoyant optimal algorithm or the analysis itself is loose. We believe the results of this paper will lead to further progress in the area of non-clairvoyant coflow scheduling.

\bibliographystyle{IEEEtran}
\bibliography{bibliography}

\end{document}